\documentclass[twocolumn,showpacs,preprintnumbers,amsmath,nofootinbib,amssymb]{revtex4}
\input psfig.sty
\def \be {\begin{equation}}
\def \ee {\end{equation}}
\def \bea {\begin{eqnarray}}
\def \eea {\end{eqnarray}}

\usepackage{graphicx}
\usepackage{dcolumn}
\usepackage{bm}

\begin{document}

\title{$\beta$-exponential inflation}

\author{J. S. Alcaniz{\footnote{e-mail: alcaniz@on.br}}}
\affiliation{Departamento de Astronomia, Observat\'orio Nacional,
20921-400 Rio de Janeiro - RJ, Brasil}

\author{F. C. Carvalho{\footnote{e-mail: fabiocc@on.br}}}
\affiliation{Departamento de Astronomia, Observat\'orio Nacional,
20921-400 Rio de Janeiro - RJ, Brasil}

\date{\today}

\begin{abstract}
An inflationary scenario driven by a slow rolling homogeneous scalar field whose potential $V(\Phi)$ is given by a generalized exponential function is discussed. Within the {\sl slow-roll} approximation we investigate some of the main predictions of the model and compare them with current data from Cosmic Microwave Background and Large-Scale Structure observations.  In particular, we show that this single scalar field model admits a wider range of solutions than do conventional exponential scenarios and predicts acceptable values of the scalar spectral index and of the tensor-to-scalar ratio for the remaining number of {\sl e-folds} lying in the interval $N = 54 \pm 7$ and energy scales of the order of Planck scale.  The running of the spectral index is briefly discussed to show that both negative and positive values are predicted by the model here proposed.

\end{abstract}

\maketitle

\section{Introduction}

Theoretical developments at the interface between high energy physics and cosmology led, about twenty five years ago, to a tremendous change in our view and understanding of the early Universe, the so-called primordial inflation \cite{inflation1} (see also \cite{revInf,book,lyth1} for a review). Similarly to the current concept of dark energy \cite{revde}, widely used nowadays to explain present cosmic acceleration, the idea of inflation, a period of rapid expansion of the cosmic scale factor in the very early Universe, became the favorite paradigm for explaining both the causal origin of structure formation and the Cosmic Microwave Background (CMB) anisotropies. From the observational side, an inflationary epoch also provides a natural explanation of why the universe is nearly flat ($\Omega_k \simeq 0$), as evidenced by the combination of the position of the first acoustic peak of the CMB power spectrum and the current value of the Hubble parameter \cite{wmap3}. 

Extending our parallel with dark energy, we must emphasize that there is also considerable freedom in modeling the field potential responsible for the primordial inflationary epoch. Several potentials, ranging from single power-laws, as the quartic $V(\Phi)\sim\lambda\Phi^4$ or the quadratic chaotic $V(\Phi)\sim m^2\Phi^2$ types, to more elaborated forms, have been largely explored in the literature \cite{Easther:2006tv}.  Another simple and interesting possibility is the one given by the  usual exponential function, i.e.,
\begin{equation} \label{potexp}
V \propto \exp{(-\lambda \Phi)}\; ,
\end{equation}
as originally investigated in Refs. \cite{exp}. Scalar fields with simple exponential potentials occur in fact quite generically in certain kinds of particle physics theories. Examples extend from supergravity and superstrings theories, as the well-studied Salam-Sezgin model \cite{halliwell}, gravitational theories with high derivative terms \cite{wett,wett1},  Kaluza-Klein theories in which extra dimensions are compactified to produce our 4-dimensional world, to many others (see \cite{wett,halliwell1} for more details). 

In this \emph{Letter}, we discuss a possible generalization for the inflaton potential (\ref{potexp}), given by
\begin{equation} \label{potexpg}
V \propto \exp_{1-\beta}{(-\lambda \Phi)}\; ,
\end{equation}
where the generalized exponential function above, defined as \cite{abramowitz}
\begin{equation} \label{potexpgdef}
\exp_{1-\beta}{({f})} = \left[1 + \beta {f} \right]^{1/\beta}\; ,
\end{equation}
\begin{eqnarray}
\mbox{for} \; \left\{
\begin{tabular}{l}
$1 + \beta {{f}} > 0$\\
\\
$\exp_{1-\beta}({f}) = 0$, otherwise,
\end{tabular}
\right.
\nonumber
\end{eqnarray}
satisfies, while ${f}$, ${g} < 0$,   the following identities:
\begin{equation}
\label{p1}
\exp_{1-\beta}\left[\ln_{1-\beta}({f}) \right] =  {f}\; \nonumber
\end{equation}
and
\begin{equation}
\label{p2}
\ln_{1-\beta}({f}) + \ln_{1-\beta}({g}) =   \ln_{1-\beta}({fg}) -  \beta \left[\ln_{1-\beta}({f}) \ln_{1-\beta}({g})\right], \nonumber
\end{equation}
where $\ln_{1-\beta}({f}) = (f^{\beta} - 1)/\beta$ is the generalized logarithmic function\footnote{A dark energy scenario derived from these generalized functions was recently discussed in Ref. \cite{prl}.}. As the real index $\beta \rightarrow 0$, all the expressions following Eq. (\ref{potexpgdef}) reproduce the usual exponential and logarithmic properties, so that the potential (\ref{potexpg}) is a direct generalization of the usual exponential function (\ref{potexp}) $\forall$ $\beta \neq 0$. For the sake of completeness, we show in Fig. (1a) the behavior of the generalized potential (\ref{potexpg}) as a function of the field $\Phi$. Note that, while $\forall$ $\beta \neq 0$ the curves show a \emph{quasi}-exponential (power-law) behavior, for $\beta = 0$ the usual potential (\ref{potexp}) is fully recovered. 

\begin{figure*}[t]
\centerline{\psfig{figure=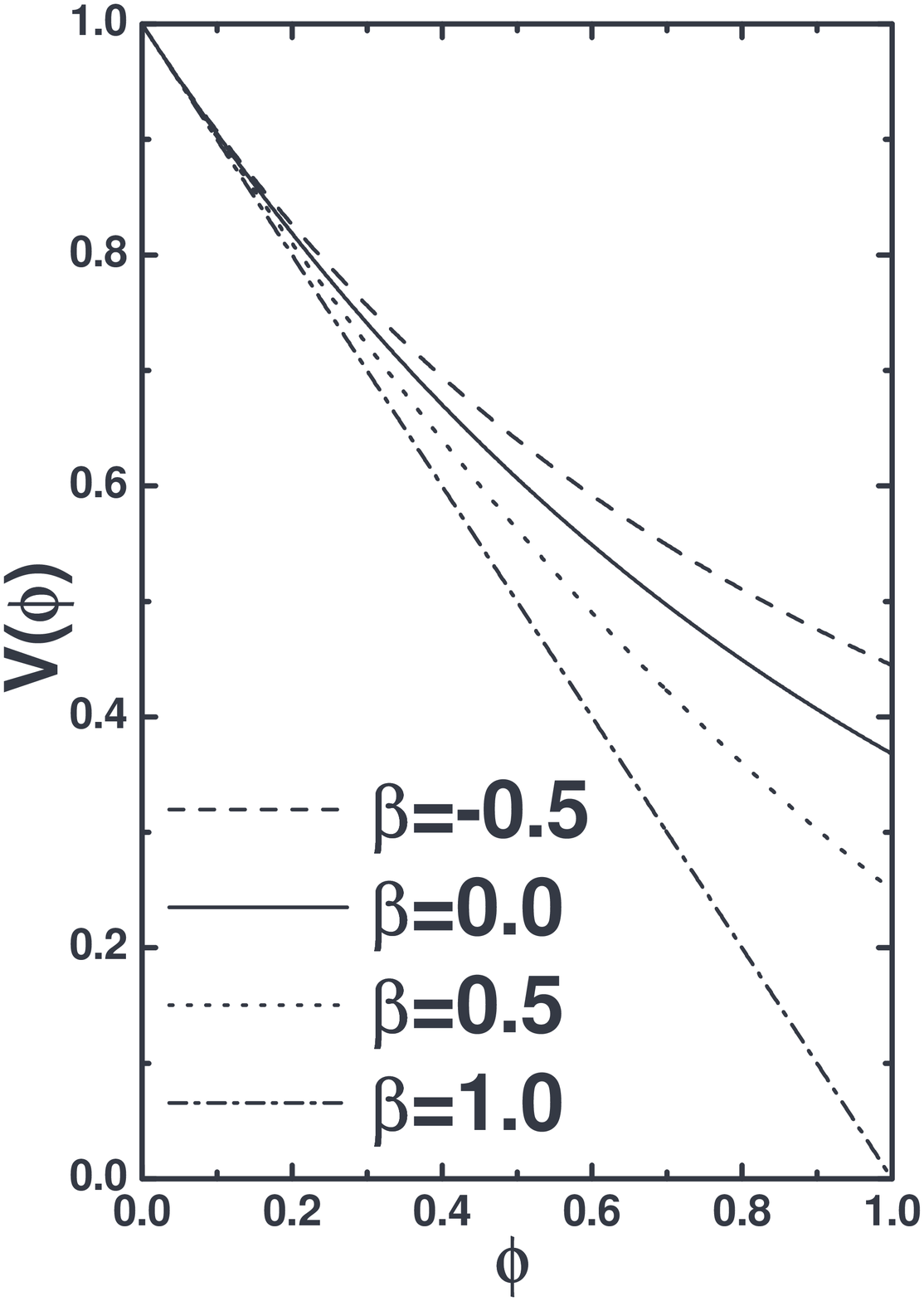,width=2.3truein,height=2.8truein,angle=0}
\psfig{figure=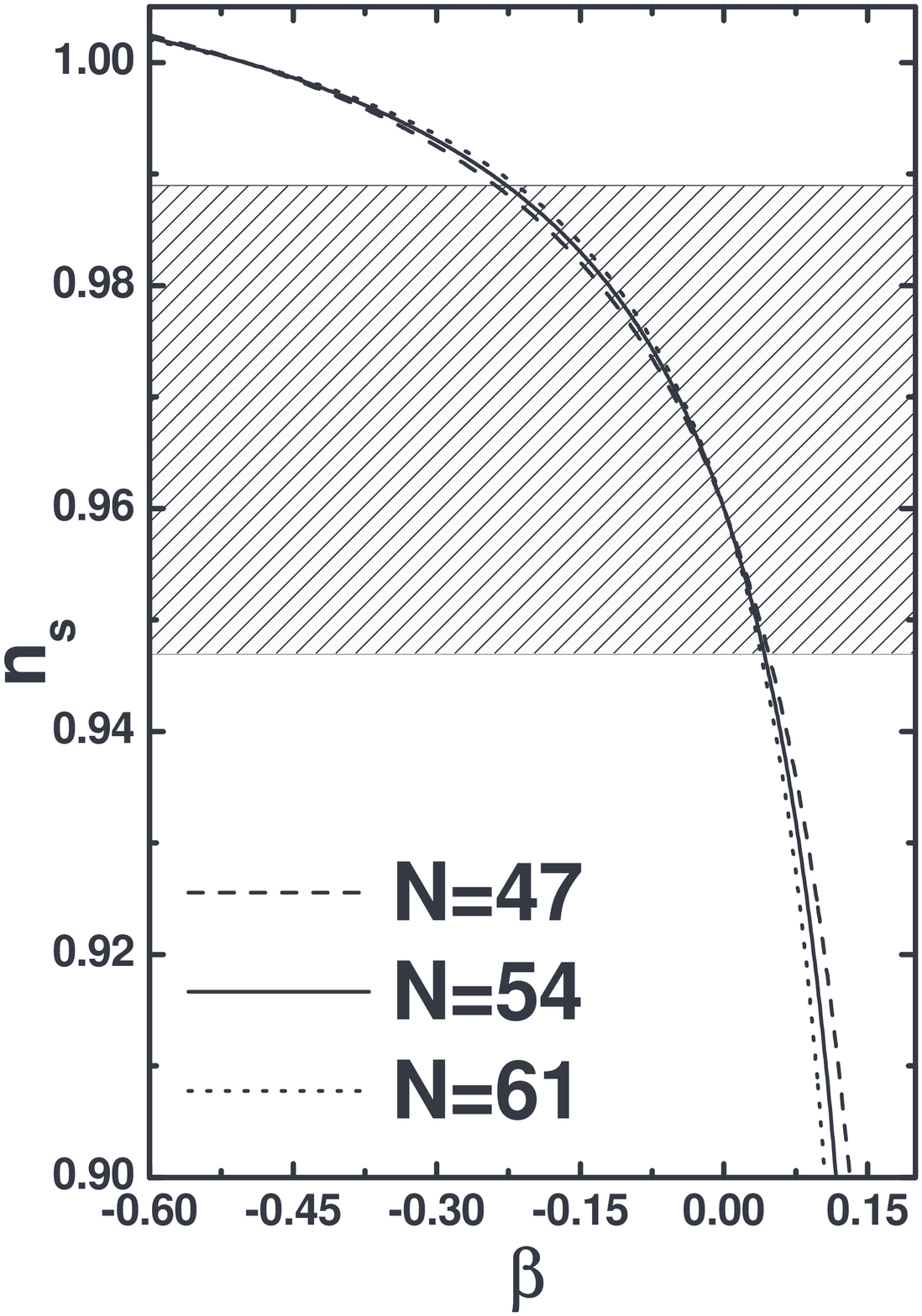,width=2.3truein,height=2.8truein,angle=0}
\psfig{figure=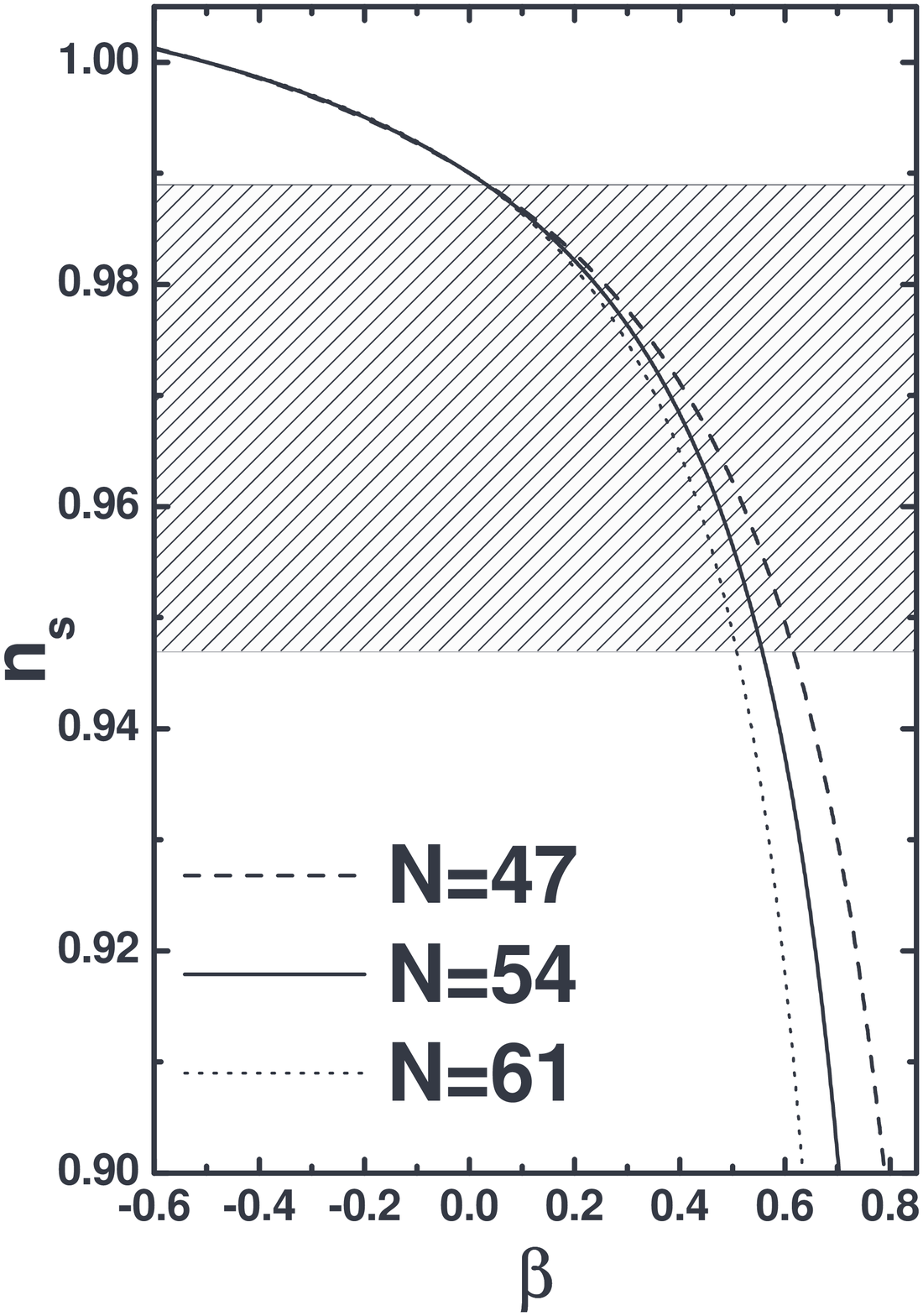,width=2.3truein,height=2.8truein,angle=0}} 
\caption{{\bf{a)}} The potential $V(\Phi)$ as a function of the field [Eq. (\ref{potexpg})] for some selected values of the parameter $\beta$. {\bf{b)}} Spectral index $n_s$ as a function of the parameter $\beta$ for selected values of the number of {\sl e-folds} ranging the interval $N = 54 \pm 7$ and $\lambda = 0.2$. Note that, for a large inteval of $\beta$ (including positive and negative values), the model's predictions are in agreement with current bounds from CMB + LSS data, i.e., $n_s = 0.967^{+0.022}_{-0.020}$ (95.4\% c.l.) \cite{tegmark} (shadowed area). {\bf{c)}} The same as in Panel (b) for $\lambda = 0.1$.}
\end{figure*}

In what follows, we analyze the inflationary scenario that arises from (\ref{potexpg}) within the slow-roll approximation. We show that this single scalar field model admits a wider range of solutions than do conventional exponential scenarios, and seems to fit the current observational constraints from CMB and Large-Scale Structure (LSS) experiments.

\section{Slow-roll parameters}

Let us first consider a single scalar field model whose action is given by  
\bea
\label{action} 
S=\frac{1}{2}\int d^4 x \sqrt{-g}\left[R - \frac{1}{2}\partial^{\mu}\Phi\partial_{\mu}\Phi - V(\Phi)\right].
\eea 
(throughout this paper we work in units where the Planck mass $m_{\rm pl} = (8\pi G)^{-1/2} = c = 1$). In this background, the stress-energy conservation equation for the field can be expressed as $\ddot\Phi + 3 H \dot\Phi + V'\left(\Phi\right) = 0$, where dots denote derivative with respect to time and primes with respect to the field $\Phi$. In the so-called {\sl slow-roll} approximation, the evolution of the field is dominated by the drag from the cosmological expansion, so that $\ddot\Phi\approx0$ or, equivalently, $3H\dot \Phi+V'\simeq 0$. With these simplifications, the {\sl slow-roll} regime can be expressed in terms of the {\sl slow-roll} parameters  $\epsilon$ 
and $\eta$, i.e., \cite{book,lyth1}
\bea
\label{epsilon}
\epsilon(\Phi) \simeq {1\over 2}\left({V'\over V}\right)^2\; ,
\eea
and
\bea	
\label{eta}
\eta(\Phi) \simeq \left[\frac{V''}{V} - \frac{1}{2}\left(\frac{V'}{V}\right)^2\right]\;,
\eea
where $H \simeq \left[V(\Phi)/3\right]^{1/2}$ is the Hubble parameter. In order to work properly, the inflationary potential must have a sufficiently small slope, so that $V'$, $V''\ll V$, which is consistent with the approximation for $\epsilon \ll 1$ and $\eta \ll 1$. 

By substituting our generalized potential (\ref{potexpg}) into the above equations, we obtain
\begin{subequations}
\begin{equation} \label{e}
\epsilon(\Phi) = \frac{\lambda^2 }{2}\frac{1}{\left[1 - \beta \lambda \Phi \right]^2}\; ,
\end{equation}
\mbox{and} 
\begin{equation}
\eta(\Phi) = \frac{\lambda^2 }{2} \frac{1-2\beta}{\left[1 - \beta \lambda \Phi \right]^2}\; ,
\end{equation}
\end{subequations}
which reduce to the usual exponential result $\epsilon \equiv \eta \equiv \rm{constant}$ in the limit $\beta \rightarrow 0$. From Eq. (\ref{e}), differently from the conventional result, one can also compute the value of the field at the end of inflation ($\Phi_e$)  by setting $\epsilon(\Phi_e) = 1$, i.e.,
\begin{equation}
\Phi_e = \frac{1}{\beta}\left[\frac{1} {\lambda} - \frac{1}{\sqrt{2}}\right]\; \quad \quad \forall \quad \beta \neq 0.
\end{equation}

In order to confront our model with current observational results we first consider the spectral index, $n_s$, and the ratio of tensor-to-scalar perturbations, $r$. In terms of the {\sl slow-roll} parameters to first order, these quantities are expressed as
\begin{subequations}
\begin{equation} \label{ns}
n_s - 1 = 2\eta - 4\epsilon = -{\lambda^2}\frac{1 + 2\beta}{\left[1 - \beta \lambda \Phi \right]^2}
\end{equation}
\mbox{and}
\begin{equation} \label{rr}
r = 16\epsilon = {8\lambda^2}\frac{1 }{\left[1 - \beta \lambda \Phi \right]^2}\; .
\end{equation}
\end{subequations}
with a direct relation between $n_s$ and $r$ given by
\begin{equation} \label{rr}
r = \frac{8(1 - n_s)}{1+2\beta}\;. 
\end{equation}
As expected, the above expression fully generalizes the well-known result for usual exponential potentials, $r = 8(1 - n_s)$ \cite {melchi}.

\begin{figure}[t]
\centerline{\psfig{figure=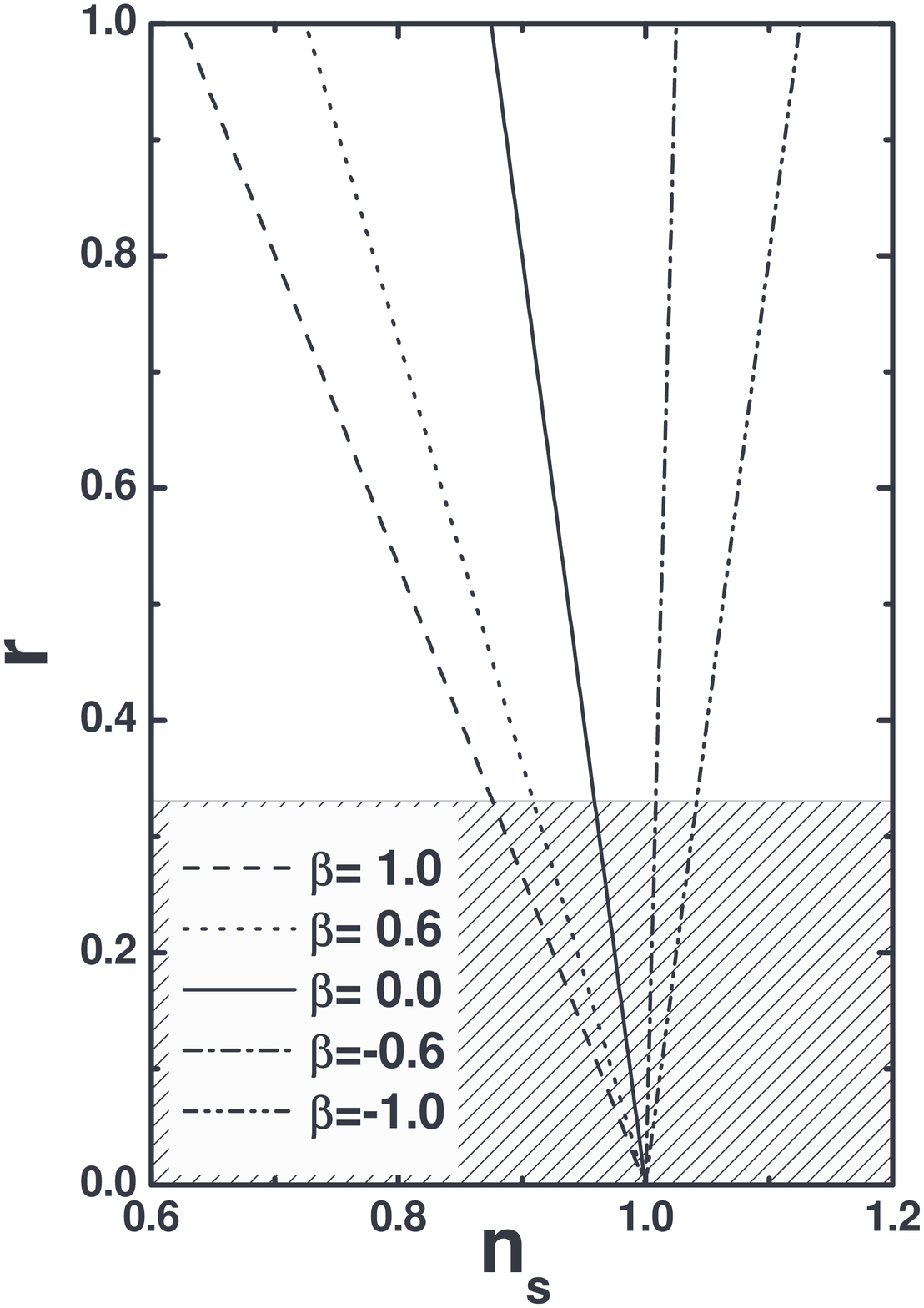,width=3.3truein,height=2.8truein}
\hskip 0.1in} 
\caption{Trajectories for different values of $\beta$ in the $n_s-r$ parametric space to first-order in {\sl slow-roll} approximation. Note that, regardless of the value of $\beta$, $n_s = 1 \Rightarrow r = 0$ [Eq. (\ref{rr})]. The shadowed area corresponds to the interval $r \lesssim 0.3$ (at 95.4\% c.l.), as given in Refs.   \cite{tegmark,melchi}.}
\end{figure}

To complete the above description, we must also calculate the number of {\sl e-folds} remaining until the end of inflation, i.e., 
\begin{equation}
N = \int_{\Phi_e}^{\Phi}{\frac{d\Phi}{\sqrt{2 \epsilon(\Phi)}}} = \frac{1}{\lambda^2}\left(\Phi_N - \frac{\beta \lambda}{2}\Phi_N^2 \right)\;.
\end{equation}
Finally, by combining the above expression with Eqs. (\ref{ns})-(\ref{rr}), we obtain the contributions of the scalar and tensor perturbations as a function of $N$, so that we can compare the model's predictions for these quantities with current observational limits. It is worth emphasizing that bounds on the gravitational wave background provide constraints on the maximum number of {\sl e-folds}, i.e., $N_{\rm{max}} \simeq 60$ \cite{n60}. In the subsequent discussions, however, we consider the interval $N = 54 \pm 7$, as well argued in Ref. \cite{lyth}.

\section{Discussion}

Figures (1b) and (1c) show the $n_s - \beta$ plane for three different values of the number of {\sl e-folds} corresponding to the range $N = 54 \pm 7$ and characteristic values of $\lambda = 0.2$ and  $0.1$, respectively. Note that, the larger the energy scale the larger the positive interval of the parameter $\beta$ that is compatible with the current bounds on $n_s$ from WMAP3 plus the Sloan Digital Sky Survey (SDSS), i.e., $n_s = 0.967^{+0.022}_{-0.020}$ (95.4\% c.l.) \cite{tegmark} (clearly an opposite dependence is also found between the remaining number of {\sl e-folds} and the positive interval for $\beta$). Note also that, for negative values of $\beta$, a scale-invariant spectrum ($n_s = 1$) is possible, which seems to be in agreement with the results of Ref. \cite{melchi} ($0.93 < n_s < 1.01$ at 95.4\% c.l.). In both figures, the former bounds on $n_s$ are represented by shadowed areas.

The $n_s - r$ plane is displayed in Figure 2 for some selected values of the parameter $\beta$. For the portion of this plane compatible with current bounds on $r$ from WMAP3 plus SDSS, i.e., $r \lesssim 0.3$ (at 95.4\% c.l.)  \cite{tegmark,melchi}, a considerable interval (which includes negative and positive values) of $\beta$ is in agreement with the current limits on $n_s$ discussed above. Similarly to the intermediate inflationary scenario of Ref. \cite{b1} (see Fig. 1 of this reference), and as expected from our Eq. (\ref{rr}), a Harrison-Zel'dovich spectrum ($n_s = 1$ and $r = 0$) is also a prediction of this scenario regardless of the value of $\beta$. In terms of the parameter $\beta$, note that both the numerator and denominator of Eq. (\ref{rr}) approaches zero as  $\beta \rightarrow -1/2$ since, from Eq. (\ref{ns}), $\beta \rightarrow -1/2 \Rightarrow n_s \rightarrow 1$.

\begin{figure}[t]
\centerline{\psfig{figure=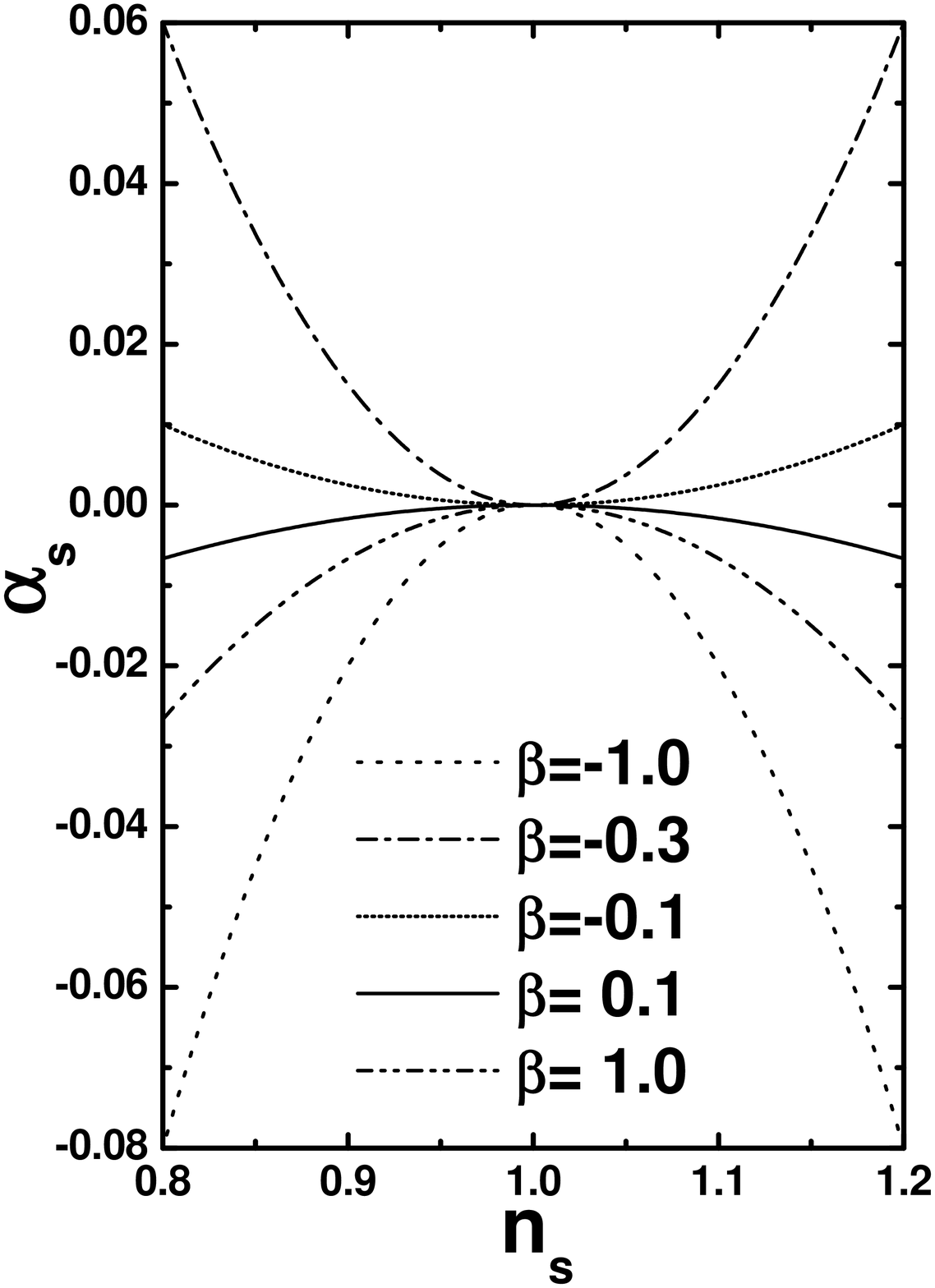,width=3.3truein,height=2.8truein}
\hskip 0.1in} 
\caption{The $\alpha_s - n_s$ plane for some selected values of the parameter $\beta$. As discussed in the text, for values of $\beta$ in the open interval ]0, -0.5[ the running is always positive while $\forall$ $\beta$ out of this interval the model's prediction is a negative running. }
\end{figure}

\subsection{Running of Spectral Index}

The running of the spectral index in inflationary models, to lowest order in slow-roll, is given by \cite{book}
\begin{equation}
\alpha_s \equiv \frac{dn_s}{d\ln k} = -8\epsilon^2 + 16\epsilon \eta - 2\xi^2\; ,
\end{equation}
where 
\be
\label{xi}
\xi^2 (\Phi) = - (2\epsilon)^{\frac{1}{2}}\frac{V'''}{V} \;.
\ee
By substituting Eqs. (\ref{epsilon}), (\ref{eta}) and (\ref{ns}) into the above expressions we obtain a relation between the spectral index and its running, i.e.,
\begin{equation}
\alpha_s = -\frac{2\beta (n_s - 1)^2}{(1 + 2\beta)}\;.
\end{equation}
Note that the above relation has both the possibilities for negative and positive running. For instance, for values of $\beta$ lying in the open negative interval $0 < \beta < -0.5$ the running is always positive (0 for $n_s = 1$), which seems to be disfavored by the WMAP3 data ($-0.02 \leq \alpha_s \leq -0.17$ at 95.4\% c.l.) but not ruled out by a joint analysis involving WMAP3 and SDSS ($0.007 \leq \alpha_s -0.13$ at 95.4\% c.l.) \cite{melchi} (see also \cite{wmap3}). For all other values of $\beta$ out of the above interval, a prediction for a negative running is found (see Figure 3). Note also that, for values of the spectral index $n_s \simeq 1$, as indicated by current observations \cite{tegmark,melchi}, all models approach $\alpha_s \simeq 0$ which, although in full agreement with the bounds above, makes a distinction between different scenarios difficult from the observational viewpoint.

\section{Final remarks}

Primordial inflation constitutes one of the best and most successful examples of physics at the interface between particle physics and cosmology, with a tremendous consequences on our view and understanding of the early Universe. In this paper, we have investigated some cosmological consequences of a new inflationary scenario driven by a generalized exponential potential of the type $V \sim \exp_{1-\beta}(-\lambda \Phi)$ [Eq.(\ref{potexpg})]. As discussed in Sec. I, this generalized potential behaves as a simple power-law for all values of $\beta \neq 0$ and is an exact exponential function for $\beta = 0$. Within the {\sl slow-roll} approximation we have calculated the main observable quantities, such as the spectral index, its running and the ratio of tensor-to-scalar perturbations and shown that, even for values of the number of {\sl e-folds} in the restrictive interval $N = 54 \pm 7$ \cite{lyth}, the predictions of the model is in good agreement with current bounds on these parameters from CMB and LSS observations, as given in Refs. \cite{melchi,tegmark}. Similarly to the intermediate inflationary scenario of Ref. \cite{b1}, it is worth mentioning that a scale-invariant spectrum ($n_s = 1$) is also prediction of the model if $r = 0$ (see Fig. 2). We emphasize that both possibilities for positive and negative values of the running of the spectral index are found for different intervals of the parameter $\beta$. Although in good agreement with current observations, we expect the next generation of experiments to be able to decide if this $\beta$-exponential potential is or not a viable possibility for describing inflation.

This work is partially supported by the Conselho Nacional de Desenvolvimento Cient\'{\i}fico e Tecnol\'{o}gico (CNPq - Brazil). JSA is also supported by Funda\c{c}\~ao de Amparo \`a Pesquisa do Estado do Rio de Janeiro (FAPERJ) No. E-26/171.251/2004.

\end{document}